# The European ALMA Regional Centre Network: A Geographically Distributed User Support Model

Evanthia Hatziminaoglou[1]
Martin Zwaan[1]
Paola Andreani[1]
Miroslav Barta[2]
Frank Bertoldi[3]
Jan Brand[4]
Frédérique Gueth[5]
Michiel Hogerheijde[6]
Matthias Maercker[7]
Marcella Massardi[4]
Stefanie Muehle[3]
Thomas Muxlow[8]
Anita Richards[8]
Peter Schilke[9]
Remo Tilanus[6]
Wouter Vlemmings[7]
José Afonso[10]
Hugo Messias[10]

[1] ESO
[2] Astronomical Institute of the Czech Academy of Sciences, Ondřejov, Czech Republic
[3] Argelander-Institut für Astronomie, University of Bonn, Germany
[4] INAF–Osservatorio di Radioastronomia, Bologna, Italy
[5] Institut de Radio Astronomie Millimétrique (IRAM), Grenoble, France
[6] University of Leiden, the Netherlands
[7] Onsala Space Observatory, Chalmers University of Technology, Sweden
[8] Jodrell Bank Centre for Astrophysics (JBCA), University of Manchester, United Kingdom
[9] I. Physikalisches Institut, Universität zu Köln, Germany
[10] Institute for Astrophysics and Space Science, Lisbon, Portugal

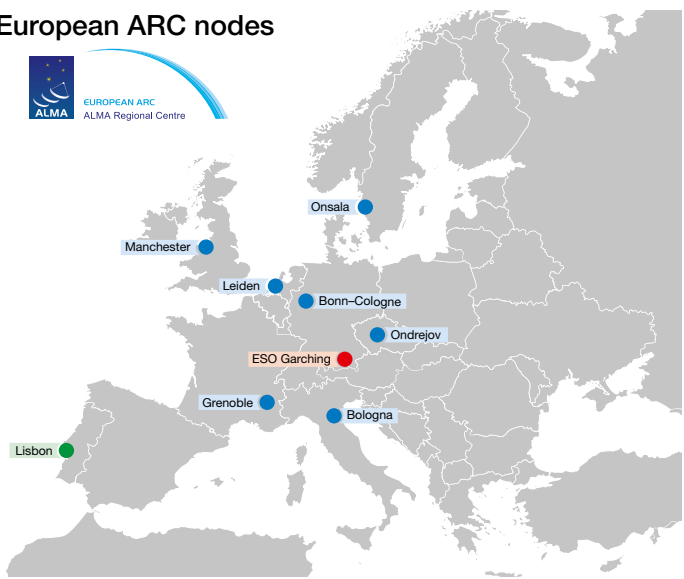

Figure 1. The locations of all the European ARC nodes are shown on a map of Europe and including the Centre of Expertise in Lisbon.

In recent years there has been a paradigm shift from centralised to geographically distributed resources. Individual entities are no longer able to host or afford the necessary expertise in-house, and, as a consequence, society increasingly relies on widespread collaborations. Although such collaborations are now the norm for scientific projects, more technical structures providing support to a distributed scientific community without direct financial or other material benefits are scarce. The network of European ALMA Regional Centre (ARC) nodes is an example of such an internationally distributed user support network. It is an organised effort to provide the European ALMA user community with uniform expert support to enable optimal usage and scientific output of the ALMA facility. The network model for the European ARC nodes is described in terms of its organisation, communication strategies and user support.

## The European ARC

A fundamental concept within the Atacama Large Millimeter/submillimeter Array (ALMA) operations model is that direct ALMA user support is not provided by the Joint ALMA Observatory[1], but by the ALMA Regional Centres. These are set up in East Asia (serving Japan and Taiwan), North America (USA, Canada) and Europe. Taiwanese users can choose to receive support from either the East Asian or North American nodes. These ARCs deliver a comprehensive ensemble of user services, ranging from general user support, helping astronomers to develop their scientific goals, all the way to assisting users with optimising data reduction and getting the most out of their ALMA observing project. Each executive (ESO, NAOJ[2], NRAO[3]) has some freedom in setting up their ARC, as long as the services agreed by all three parties are delivered.

At an early ALMA Community Day in Garching in 2002 (Shaver & van Dishoeck, 2002), the attendees voted for a European ARC model with a central, coordinating node at ESO and a number of smaller nodes in the community. The ESO Scientific Technical Committee (STC) and ESO Council approved a Call for Statements of Interest to establish local ARC nodes, which was sent out in 2004, and resulted in responses from six institutes in Europe. The initial model of the European ARC was described in Andreani & Zwaan (2006). A Memorandum of Understanding between ESO and the six ARC nodes was signed in 2008, marking the official start of the European (EU) ARC network.

Today, in 2015, the EU ARC network still operates with the same six founding members: Italian (located in Bologna), German (in Bonn/Cologne), IRAM (in Grenoble), Dutch (Allegro; in Leiden), British (in Manchester) and Nordic countries (based in Onsala). In addition, an ARC node in Ondrejov (Czech Republic) was created in 2009. The most recent extension of the European ARC network is the Portuguese Centre of Expertise (CoE), which has been operating in Lisbon since early summer 2014. A CoE is a temporary status through which potential new nodes may have to transition while they build up the required expertise and user base. Figure 1 shows the geographical distribution of these nodes and the CoE.

For convenience, the nodes are named after the communities that they primarily serve. They also serve users from outside these communities who do not have a "local" node, or simply wish to take advantage of the expertise and experience available at a particular node. Each node has the mandate to primarily serve their local community. The staff at each ARC node are scientists covering a range of expertise in interferometry, (sub-)millimetre observing and ALMA data reduction and interpretation. The ARC nodes are based at locations in Europe that have a long history in radio and/or millimetre observing and data analysis. The ARC network therefore makes optimal use of the immense knowledge and expertise that exists in Europe, without having to build this up at ESO alone.

The central node, usually referred as "the ARC", is located at ESO Headquarters in Garching. The ARC is set up as a department within the European ALMA Support Centre (EASC), a cross-divisional structure within ESO, which, in addition to the ARC, also houses the ALMA Technical Support Group, the ALMA Science Programme and ALMA Outreach.

### ARC network management and coordination concepts

An interesting question to consider is why the European ARC network functions, if there are no contractual obligations between ESO or ALMA and the ARC nodes. ESO plays a managing and coordinating role, but the ARC nodes are autonomous in the way they set up their local structures. They have complete freedom over the number of staff they hire, the expertise areas on which they wish to concentrate, and so on. Furthermore, ESO does not provide financial support for the ARC nodes. The ARC nodes seek their own funding through a range of national schemes. Therefore, funding horizons may vary from one to five years between the different nodes.

Rather than having formal or legal agreements, the ARC network partnership is based on the Memorandum of Understanding, but above all on trust and collegiality. This collaboration, that at first sight may seem vulnerable, functions well because of its symbiotic nature: through the ARC nodes ESO is able to provide the European ALMA users with services that far exceed what would have been possible without the network, while the ARC nodes are able to build up expertise in ALMA operations and technical developments and secure funding for their institutes.

Coordinating and managing the ARC network requires a dedicated effort from ESO and each of the ARC nodes. Three ESO astronomers are tasked with bringing the diverse elements of the complex ARC structure into one efficient working model, referred to as the EU ARC manager, liaison and coordinator. The European ARC manager is responsible for the successful operation of the European ARC. The ARC manager is also member of the ALMA Science Operations Team and the EASC management team, and has a number of different roles, including being responsible for running the ARC at ESO. In addition, the ARC manager provides a crucial link between the ALMA Director's Office and ESO Management, and is directly accountable to the ALMA Director for Board-approved operational deliverables. The ARC node liaison is responsible for matters related to the nodes, and oversees the provision of support and overall coordination, working closely with the ARC manager. Finally, the ARC node coordinator acts as the contact person for all ARC staff, and is responsible for maintaining communications.

This team of three people is referred to as the ARC network management team and together they are responsible for the smooth functioning of the ARC network. In addition to this team, at each of the ARC nodes one representative is responsible for the communication with the network and the local coordination of their node. The ARC node representatives, together with the ARC network management team, are collectively referred to as the ARC Coordinating Committee, or ACC. Important decisions within the ARC network are taken by the ACC through consensus decision-making.

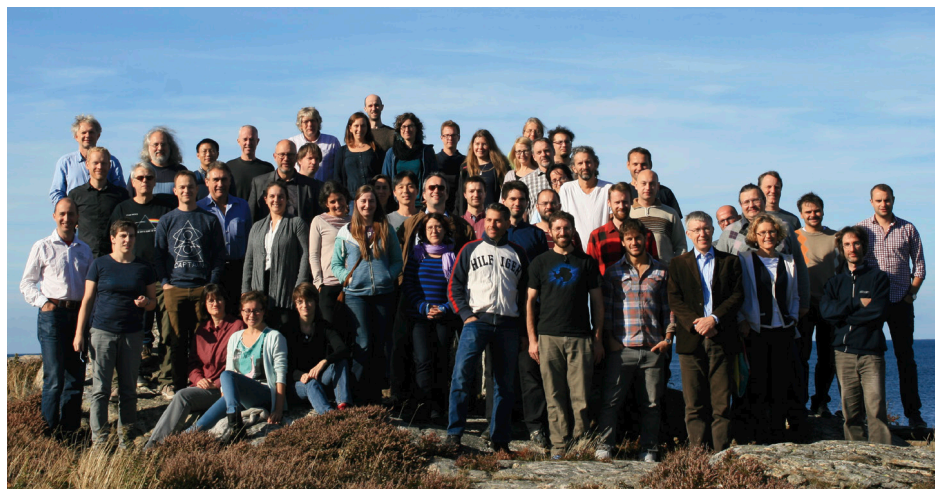

Figure 2. Many of the staff of the nodes of the European ARC pictured during the most recent all-hands meeting in Smögen, Sweden, 28–30 September 2015.

### Communication scheme at the heart of a uniform user experience

The dispersal of ALMA user support in Europe to a number of geographically separated institutions (Figure 1) often makes it a challenge to ensure the dissemination of information in a steady and effective way. In addition, the geographical separation of the ARC nodes from the ALMA Observatory implies that activities in the European ARC nodes run the risk of not being optimally aligned with ALMA developments carried out elsewhere, or that the nodes might feel detached from the observatory, and not an integral part of the ALMA project. Information flow in all relevant directions is therefore essential and, in turn, results in a high level of homogeneous user support across the network, while satisfying the specific needs of the local communities.

A number of interfaces between the ARC at ESO and the nodes are in place to enable and promote communication. The ARC TWiki, hosted and maintained by ESO, is a document repository and a means to homogeneously disseminate and store information within the network. All European ARC staff have access to it. Monthly telecons/videocons, organised and chaired by ESO and attended by the ACC, facilitate dialogue and coordination among the nodes, keep people up to speed with the latest ALMA developments and network news, and help to identify and rapidly react to any problems.

Face-to-face meetings are still the most effective ways of communicating and diffusing information. Once a year an "all-hands" meeting, open to all European ARC staff, is organised and the venue rotates through each of the nodes and ESO. This year's meeting was held in Sweden (see Figure 2). This is a superb opportunity to go through the ARC activities over the past year, discuss ALMA operations in general and plan for the future. In a geographically dispersed network these yearly meetings are essential. ARC staff are informed of the latest ALMA news and developments and get a chance to discuss issues and concerns openly or in smaller groups. But most importantly, staff get to know each other and strengthen working relationships that could otherwise remain formal and impersonal due to the distance.

A second type of yearly face-to-face meeting takes place among ACC members, including the ARC network management team, with the participation of one additional member from each ARC node. The purpose of the meeting is to discuss high-level operational requirements, issues that touch upon the ARC network organisation and communication with the ALMA project, and the short- and medium-term evolution of the network and its functions. The venue of this meeting also rotates among the various nodes and ESO. The ARC network management team also visits the nodes at a frequency set by the individual needs of each node. This gives all the involved parties the opportunity to concentrate on the services provided by each of them, to discuss whether they are satisfied with the current arrangements and to establish whether any improvements could be made. Finally, European ARC staff regularly meet during ALMA training and scientific events as well as on a number of other formal or informal occasions.

Information related to the use of the ALMA facility has to be shared with the scientific community. The European ARC webpages[4] contain high-level information regarding the activities and services of the network as well as links to the local web pages of the nodes, and is a good starting point for users seeking specific information regarding the network. The central point of contact for all ALMA users is the ALMA Helpdesk[5], which is available to all registered users from the ALMA Science Portal[6] in one uniform users' view. There is one Helpdesk for all global ALMA users, but queries from European users are automatically directed to and addressed by the European ARC. As part of the common user experience, the ALMA Science Portal also shares one unified view for all users across the globe, but has three mirror sites, and users are automatically redirected according to the location from which they are connecting. The only difference between the three ARC mirror sites is the column hosting the Local News, specific to each ARC. Important ALMA information also reaches the European ALMA users via their regional node newsletters, as well as via the newly established European ARC newsletter, which will be issued every three months.

The experience after seven years of ARC network collaboration, backed up by research into the performance of geographically dispersed teams (Huang, 2012), shows that communication is often the Achilles heel of the collaboration: team members simply cannot interact as easily as in a centralised team. Good communication is therefore essential in keeping a distributed network functional. It should have low thresholds, should use all media and formats that people find convenient (telephone, video connections, face-to-face meetings), and must be regular.

The European ARC user support model

The staff at ESO and the ARC nodes work together to provide optimal support to users during the complete lifetime of a project, from proposal preparation, through the creation of the scheduling blocks (SBs), to delivery of the calibrated

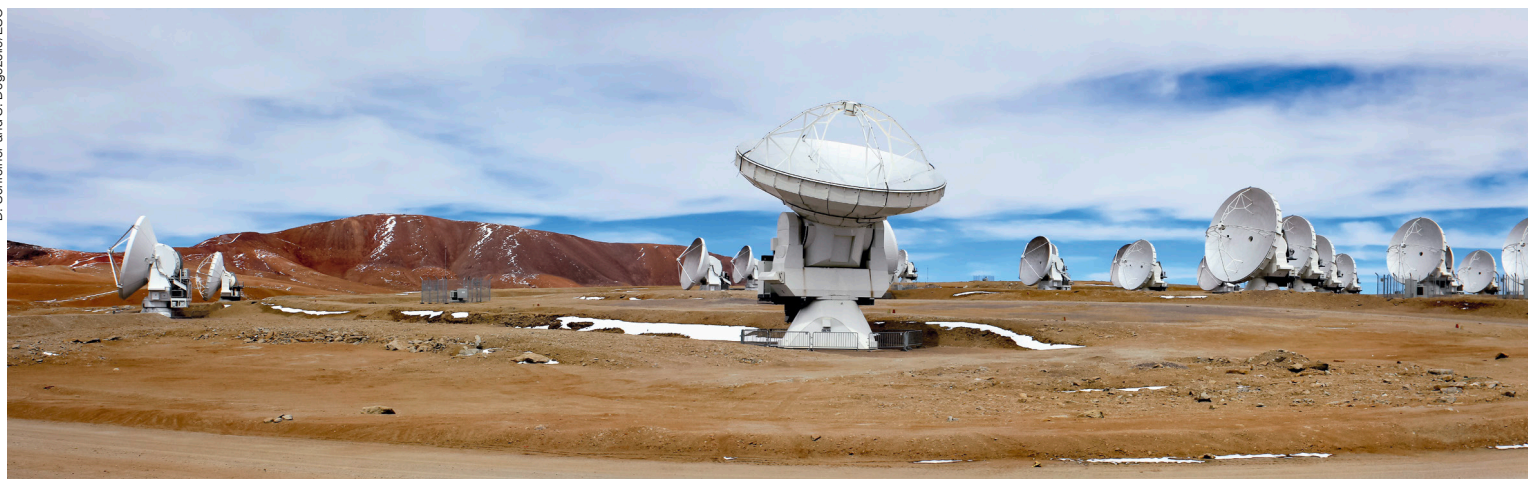

D. Schreiner and S. Degezelle/ESO

science products to the users, and, if required, to additional data reduction support. Personalised face-to-face support is the biggest success of the European ARC network and the *raison d'être* of the ARC nodes. In what follows we describe this model in more detail.

Users with newly approved observing projects are notified by email and assigned to an ARC Contact Scientist (CS), usually from the local ARC node. The ARC CS works with the PI and the Phase II team at ESO to create the SBs. The PI is notified when the SBs are ready for approval and the CS draws their attention to (and, if necessary, explains) any significant issues so that the PI can check and finally approve the SBs. Once all Phase II material is approved by the PI and technically validated, it is submitted to the ALMA Archive ready for scheduling and eventual execution. PIs can then monitor the progress of their projects in terms of observations and data reduction via dedicated tools.

After the successful execution of the project and subsequent data delivery, the PI can ask for face-to-face support for regular or advanced data reduction. Face-to-face support can also be requested for the purposes of proposal preparation or archival research. A complete, synchronised mirror of the ALMA Archive is kept at ESO[7], and is a valuable resource for data mining. ALMA datacubes are particularly complex and the scientific information which they carry can be used for scientific goals other than the ones in the original proposal. In order to ensure that the archive as well as ALMA itself are exploited to their full potential, the nodes also provide support for users wishing to carry out archival research.

Requests for face-to-face support are usually received via the ALMA Helpdesk. Requests made directly to the nodes are likewise transmitted to the Helpdesk. Once a node is made aware of a visit request, support staff at this node are responsible for arranging the details of the visit. Visits are usually made to a user's regional node as national funding bodies expect this. However, if a certain area of expertise is required, a visit to a node other than the local node can be arranged, if this will provide the user with the best possible support. Each visitor is assigned a single member of staff at that ARC node for all aspects of support.

After a visit, the user is encouraged to submit (anonymously, if so desired) feedback via a web form. Feedback is also required from the node support staff in order to discuss any technical aspects of the support or successful data reduction of technically challenging projects that other nodes might encounter in the future. At the time of writing, 89 formal face-to-face requests have been recorded in the Helpdesk from Cycle 0 to Cycle 2, with the majority asking for support with data reduction. This number does not include less formal visits involving users at close proximity to a node, who meet with ARC node staff in order to get extra help with their data interpretation, future proposals, etc.

### Diversity: A strength within a weakness

To all intents and purposes, the network for the European ARC nodes already operates as a virtual organisation (Travica, 1997) obeying two structural conditions: geographical dispersion of the organisational units (nodes) and the use of information technology to support work and communications. A virtual organisation resembles a traditional organisation in its inputs and outputs. It differs in the way in which it adds value during the journey between these inputs and outputs (Economist, 2009). The coordination of such a structure implies an understanding of the cultural and interpersonal differences, and its successful operation relies on smooth communications, a common aim and a shared sense of responsibility, all the while respecting the distinctiveness of each individual organisational unit. Subsequently, the challenges such a structure faces are many, starting from establishing a seamless information flow, to ensuring that all decisions are understood and executed in exactly the same manner by all parties involved, as well as building a common sense of purpose, sharing risks and responsibilities and finally jointly enjoying and benefitting from a successful outcome.

The weakness as well as the strength of the ARC network lies in its heterogeneous structure: nodes have different organisations, funding schemes, community size and expertise, and — perhaps more importantly — mandates from their funding agencies. This means that they constantly have to keep a balance among, sometimes conflicting, requirements coming from their users, funding bodies, the ARC network and the ALMA project itself. At the same time, such differences foster the development of new ideas and new concepts of user support, impose flexibility, and broaden the scope of the network.

Figure 3. A panoramic view of ALMA antennas on the Chajnantor Plateau.

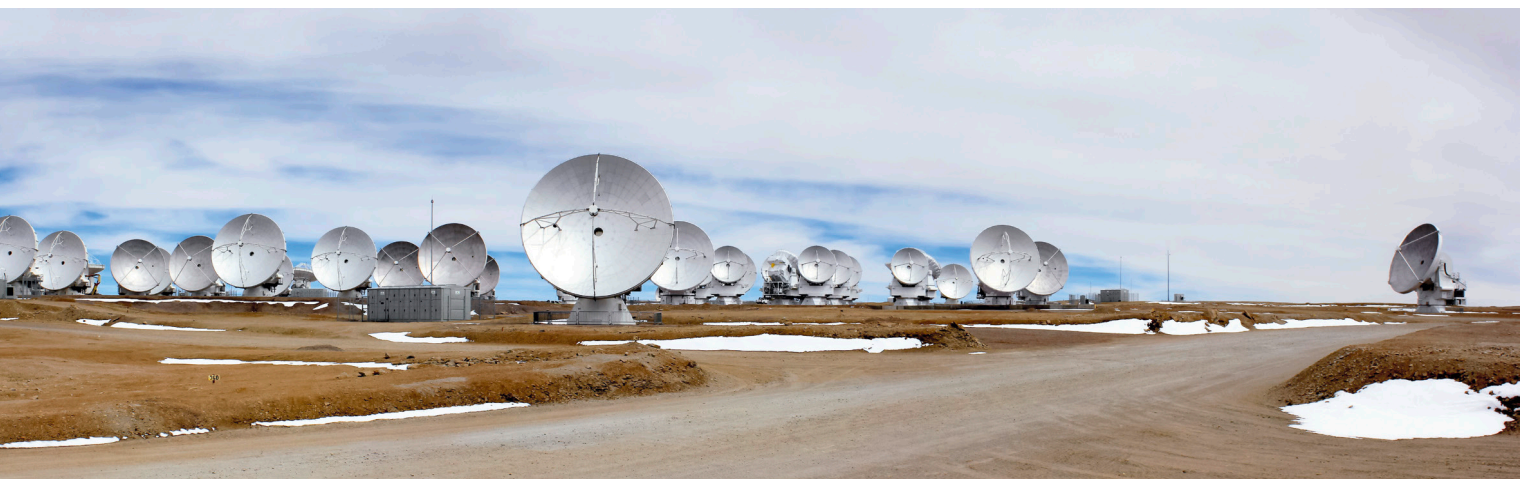

In order to keep the ARC network a lively entity, with the flexibility to respond to external factors, such as varying funding situations, changing user needs, and unplanned tasks, it is necessary to identify a number of items that are crucial for the ARC network to maintain:

Face-to-face user support. In this area, the European ARC is unique in that many more users in Europe make use of this service compared to the other executives, and the service is valued extremely highly, as shown by feedback[8]. A distributed support network that is geographically close to the users and makes optimal use of the large body of millimetre/submillimetre interferometric expertise in Europe implies a clear competitive advantage for European astronomers.

Unique scientific or technical expertise. Within the European ARC network there are several expertise areas that are unique within the ALMA project, including Solar observation, polarimetry, high frequency and long-baseline phase correction and array combination. European ARC nodes have contributed particularly extensively to ALMA commissioning in these areas, and are in an ideal position to provide optimal specialised user support.

Enhanced services. The European ARC network has a long tradition of providing software packages to ALMA and to the user community. These range from packages that assist the Observatory (e.g., the data packager or the water vapour radiometer correction software), to user support and science analysis software, such as radiative transfer codes.

General ALMA development. To maintain its role as a premier observational facility in the coming decades, ALMA maintains a healthy long-term development programme. Advanced observing modes and capabilities are rolled into the array operational environment to respond to the evolving challenges of astrophysical research. Some examples from the early years of ALMA development are the new Band 5 receivers and millimetre-wave very long baseline interferometry (VLBI) capabilities. Many of the ARC nodes are involved in the development of new ALMA capabilities, their implementation at the observatory and the future user support needs.

The network for the ARC nodes closely coordinates the development of these areas of expertise to ensure global coverage, avoid overlap and stimulate inter-node collaborations.

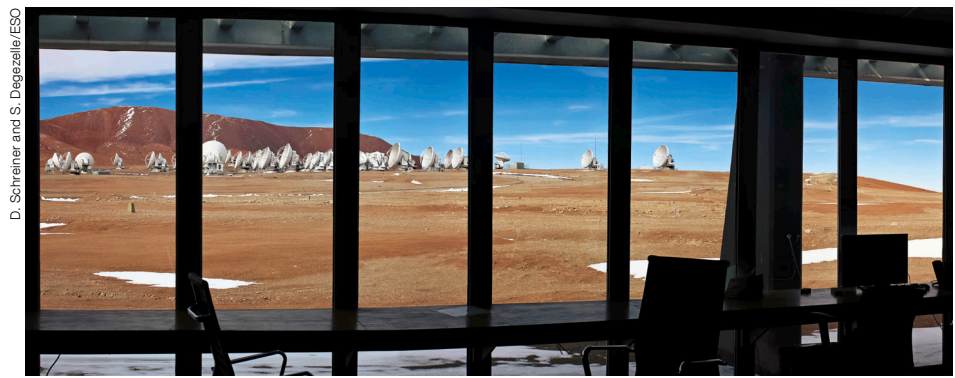

Figure 4. The view out of the ALMA Array Operations Site (AOS) Technical Building, that houses the ALMA correlator.

## The evolution of the ARC network

The ARC nodes have been a great asset in attracting young people who aspire to get close to the operational side of the ALMA project and pursue a career in science. Staff at ARC nodes are scientists who are trained in operating the facility, understand its technical aspects, gain valuable technical skills and provide user support to facilitate the best science return from the facility. However, one of the most challenging issues that the network continually faces is to retain the expertise acquired over the years, as about half of the people in the network are on fixed-term contracts. Having to import expertise is a healthy way of renewing the "population" of the network. It does however require a robust training scheme that ensures skill transfer among both staff and nodes, on timescales that do not interfere with the smooth functioning of the network and the services provided. To this aim, training workshops are regularly organised inside the network, covering both procedural (e.g., Contact Scientist training) and operational (e.g., data reduction, use of calibration pipeline, etc.) skills.

As ALMA observations become more standardised, and experience in the community with ALMA data grows, user support at the nodes will shift from direct assistance with technical and/or practical aspects, to pushing the ALMA capabilities to the limit, by supporting large, data-intensive projects and enabling advanced scientific analysis. User support provided by the ARC nodes may, therefore, progressively acquire a more science-oriented character, while maintaining the basic support necessary to new users or use of new ALMA capabilities. To accommodate the needs of an increasingly demanding ALMA user community, the ARC nodes work towards strengthening their areas of expertise, such as polarimetry, (millimetre-wave) VLBI, Solar observations, or the enhancements of synergies between ALMA and the other facilities that each of the ARC nodes is actively supporting.

Furthermore, driven by the needs of their communities, already established mandates or the requests from their funding agencies, some ARC nodes already act as, or are developing towards, general support centres. Such centres provide support for a multitude of instruments and facilities operating in the (sub)millimetre and radio regimes, maximising the synergies between ALMA and various other (European and non-European) facilities.

The European ARC network is an ever-evolving structure, due to both its organisation and the nature of the support it is providing. It is therefore not only the expertise and type of support that may change, but the number of nodes may evolve with time, as well. New nodes (or CoEs) may join the network just as old nodes might leave as a result of the changing needs of the user base or decisions by the funding agencies.

Concluding remarks

In January 2015, the ARC network was reviewed by an external committee. Receiving an independent assessment of the key aspects of the network turned out to be an extremely useful exercise. It clearly brought up the strengths of the network and the reasons behind the network's success. At the same time, it motivated an honest discussion on its vulnerabilities that has helped to establish a course of action that takes into account the evolving needs of both the network and the European ALMA user community.

Undoubtedly, the reasons behind the continuing success of the network are the mutual benefits: the nodes have become an integral part of the ALMA project, they enhance their expertise and provide a service greatly appreciated by their communities and their funding agencies. While ALMA has been extensively advertised through well-organised community events, users appreciate direct help from well-known experts close to their home institutes. At the same time, the European ARC gains from the large pool of expert staff working at the nodes, securing user support at a level that would have been impossible without them. ARC node staff are also actively using the facility in order to pursue their own scientific and technical interests. In addition to enhancing their expertise, this connection increases their motivation and involvement and makes them effective ambassadors of the ALMA project.

In order to ensure the continuation of the seamless functioning of the network,

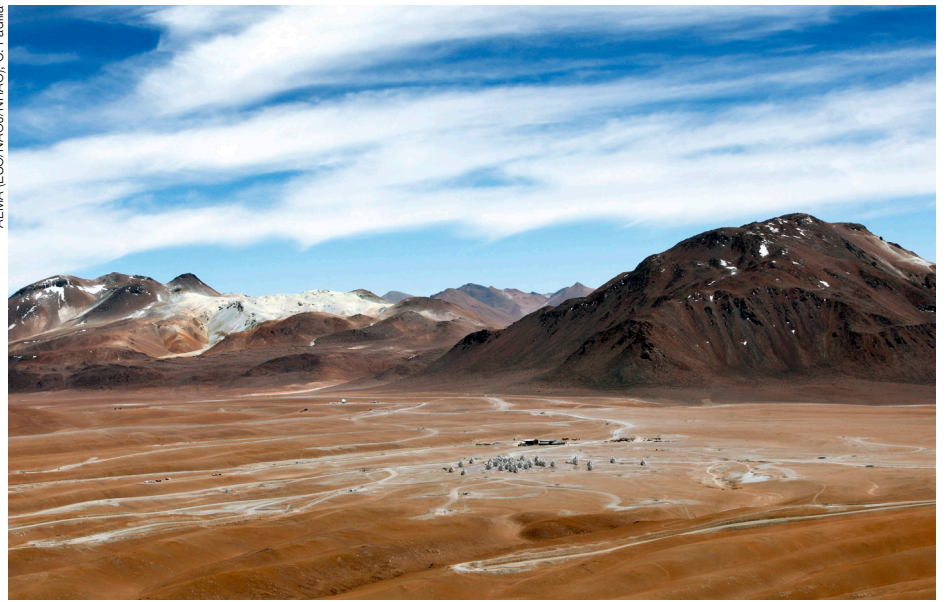

efforts have to be made towards maintaining the goodwill that has developed, primarily as a result of the good communications, and also in securing the recognition of the skills and the professionalism of the ARC node staff, and their subsequent level of engagement. A good way to maintain the latter is by establishing their high level of involvement in the commissioning efforts of new capabilities, thus bringing visibility as well as added knowledge.

Flexibility is an essential quality of the network. The quick reaction of the ARC network to the huge workload brought about by the, originally unaccounted, demand for quality assurance is an example demonstrating that the network is flexible enough to deal with complications when they arise. However, nothing should be taken for granted, given the differences in the organisational and funding schemes behind each individual node. The balance that the network has achieved to date needs to be maintained. The ARC at ESO has neither the duty nor the wish to control the detailed management of the nodes. The nodes will continue to contribute the amount of effort that is beneficial to them, taking into account their own needs and those of the community that they support. Ultimately, the capacity of the network to develop and deliver in the long term will critically depend on its ability to function as a single distributed team.

Figure 5. A view across the Chajnantor Plateau.

Acknowledgements

This project requires a huge amount of work and dedication. Its implementation and continuous success would not have been possible without the hard work of all former and present members of the European ARC network. We also thank the different funding agencies that support the individual ARC nodes, and together make the existence of this network possible.

Links

[1] Joint ALMA Observatory: http://www.almaobservatory.org
[2] National Astronomical Observatory Japan: http://www.nao.ac.jp/en/
[3] US National Radio Astronomy Observatory: http://nrao.edu/
[4] European ARC web pages: http://www.eso.org/sci/facilities/alma/arc.html
[5] ALMA helpdesk: http://help.almascience.org
[6] ALMA Science Portal: http://www.almascience.org
[7] ALMA archive at ESO: https://almascience.eso.org/alma-data/archive
[8] ALMA Cycle 3 User Survey: https://almascience.eso.org/news/alma-cycle-3-user-survey-3